\documentclass{aa}

\usepackage{amsmath}
\usepackage{graphicx}
\usepackage{txfonts}
\usepackage{hyperref}
\usepackage{multirow}

\DeclareTextFontCommand{\textbf}{}

\begin{document}

\title{Multi point analysis of coronal mass ejection flux ropes using combined data from Solar Orbiter, BepiColombo and Wind}

    \author{Andreas J. Weiss \inst{1, 2}
           \and Christian M\"ostl \inst{1}
           \and E.E.~Davies \inst{4}
           \and Tanja Amerstorfer \inst{1}
           \and Maike Bauer \inst{1, 2}
           \and J\"urgen Hinterreiter \inst{1, 2}
           \and Martin A. Reiss \inst{1}
           \and Rachel L. Bailey \inst{3}
           \and T.S.~Horbury \inst{4}
           \and H.~O'Brien \inst{4}
           \and V.~Evans \inst{4}
           \and V.~Angelini \inst{4}
           \and D.~Heyner \inst{5}
           \and I.~Richter \inst{5}
           \and H-U.~Auster \inst{5}
           \and W.~Magnes \inst{1}
           \and D.~Fischer \inst{1}
           \and W.~Baumjohann \inst{1}
          }

    \institute{Space Research Institute, Austrian Academy of Sciences, Schmiedlstraße 6, 8042 Graz, Austria\\ \email{andreas.weiss@oeaw.ac.at}
              \and Institute of Physics, University of Graz, Universit\"atsplatz 5, 8010 Graz, Austria
              \and Zentralanstalt f\"ur Meteorologie und Geodynamik, Hohe Warte 38, 1190 Vienna, Austria
              \and Department of Physics, Imperial College London, London, UK 
              \and Technical University of Braunschweig, Braunschweig, Germany
             }
    \date {Received: 29 March 2021 | Revised: 17 May 2021 | Accepted: -}
    \abstract
    {
    \textbf{The recent launch of Solar Orbiter and the flyby of BepiColombo opened a brief window in which these two spacecraft, and existing spacecraft at L1, were positioned in a constellation that allows for the detailed sampling of any Earth-directed coronal mass ejection (CME)}. Fortunately, two such events occurred during this time period with in situ detections of an interplanetary coronal mass ejection (ICME) by Solar Orbiter on the 2020 April 19 and 2020 May 28. These two events were subsequently also observed in situ by BepiColombo and Wind around a day later.
    }{
    We attempt to reconstruct the observed in situ magnetic field measurements for all three spacecraft simultaneously using an empirical magnetic flux rope model. This allows us to test the validity of our flux rope model on a larger and more global scale. It additionally allows for cross-validation of the analysis with different spacecraft combinations. Finally, we can also compare the results from the in situ modeling to remote observations obtained from the STEREO-A heliospheric imagers which were able to capture the interplanetary evolution of the coronal mass ejections.
    }{
    We make use of the 3D coronal rope ejection model (3DCORE) in order to simulate the ICME evolution and reconstruct the measured flux rope signatures at the spacecraft positions. For this purpose, we adapt a previously developed approximate Bayesian Computation sequential Monte-Carlo (ABC-SMC) fitting algorithm for the application to multi point scenarios. This approach not only allows us to find global solutions, within the limits of our model, \textbf{but also naturally generate error estimates on the model parameters and detects potential ambiguities}.
    }
    {We show that we are able to generally reconstruct the flux ropes signatures at three different spacecraft positions simultaneously using our model in combination with the flux rope fitting algorithm. For the well-behaved April 19 ICME our approach works very well while only showing minor deficiencies. The May 28 ICME, on the other hand, \textbf{shows the limitations of our approach for \textbf{less clear} ICME measurements or strongly deformed shapes}. Unfortunately, the usage of multi point observations for these events does not appear to solve inherent issues, such as the estimation of the magnetic field twist or flux rope aspect-ratios due to the specific constellation of the spacecraft positions, which all lie near the ecliptic plane. As our general approach can be used for any fast forward simulation based model we give a blueprint for future studies using more advanced ICME models.}
    {}
    \titlerunning{Multi point analysis of coronal mass ejection flux ropes}
    \maketitle

\section{Introduction} \label{sec:intro}

Coronal mass ejections (CMEs) are a highly energetic process in which a large amount of magnetized plasma is ejected from the Sun. Interplanetary CMEs (ICMEs), which can be observed and identified using heliospheric imagers, propagate far into the heliosphere and arrive at the inner planets within a few days \citep{Moestl_2009A, Rouillard_2011}. On Earth, they are responsible for the most extreme geomagnetic storms that can occur and are therefore the strongest possible driver for the Earth's space weather system \citep{Gosling_1991, Farrugia_1993, Gonzalez_1999, Schwenn_2006}. As such, they represent one of the most extreme events that must be taken into account when relying on modern applications or equipment that are affected by the space weather environment \citep[e.g.][]{Fuller_Rowell_1994, Bolduc_2002, Pulkkinen_2017}.

While the geomagnetic effectiveness of an ICME arriving at Earth relies on multiple factors, the two most important indicators are the speed and the north-south magnetic field component $B_z$. Given a strong southward orientation of the magnetic field, a ICME can facilitate the injection of energy into the magnetosphere leading to the aforementioned geomagnetic events \citep{Burton_1975, Gonzalez_1994}. The in situ magnetic field signatures of ICMEs often exhibit similar structure with a frontal shock and sheath region that is followed by a highly structured rotating magnetic field \citep{Burlaga_1981, Klein_1982, Kilpua_2017}. This rotating magnetic field, which is commonly referred to as the magnetic obstacle, can potentially manifest itself a strongly south orientated magnetic field over the duration of multiple hours. As such, the understanding of the structure of these magnetic obstacles is key to linking ICMEs with the resulting geomagnetic effects \citep{Bothmer_1998}.

\textbf{The magnetic obstacles are often described as magnetic flux ropes (MFR) structures \citep{Marubashi_1986, Burlaga_1988, Lepping_1990}. While alternative pictures exist, based on spheromak models \citep{Vandas_2017, Singh_2020}}, the most common depiction uses some form of a bent flux tube that remains connected to the solar surface while propagating away from the Sun \citep{Dasso_2005}. A variety of different successful magnetic flux rope models have been developed for the purpose of describing the in situ signatures of ICMEs which generally assume a rigid cylindrical or toroidal geometry. They are furthermore divided into two general classes, those based on uniform-twist fields \citep{Gold_1960, Hu_2015, Vandas_2017} or constant $\alpha$ fields \citep{Lundquist_1950, Lepping_1990, Hidalgo_2002, Nieves_Chinchilla_2018}.

Along with the wide variety of descriptions for the magnetic field, the global shape of ICMEs also remains largely hidden. While the initial structure can be partially observed due to Thomson scattering close to the Sun \citep{Cremades_2004, Thernisien_2006}, the in situ magnetic field measurements represent a strong projection of the internal magnetic field that is intertwined with the global or local geometry. This can make it exceptionally difficult to draw conclusions on the shape using just single in situ magnetic field measurements. It is broadly understood that there are various processes that influence the shape and evolution of ICMEs during their propagation through the heliosphere \citep{Manchester_2017}. This includes rotations or deflections close to the Sun \citep{Lugaz_2012, Kay_2015} and interactions with the ambient solar wind \citep{Riley_2004A, Liu_2006, Demoulin_2009}. More complex interactions are also possible such as CME-CME collisions \citep{Moestl_2012, Lugaz_2017}. 

More recently there has been a focus on the study of multi point events, i.e. ICMEs that are observed in situ by multiple spacecraft \citep{Moestl_2009B, Kilpua_2009, Good_2019, Davies_2020, Salman_2020} in an attempt to gain a better global understanding of ICME structure and their evolution. Unfortunately, such events are inherently rare due to the limited number of spacecraft with on board magnetometers in different solar orbits within similar time frames.  The successful modeling of these multi point events invariably requires global ICME models, that include magnetic fields and ICME evolution, which goes beyond the capability of analytical models with rigid cylindrical or toroidal shapes. \textbf{While it may be tempting to use large scale MHD models \citep[e.g.][]{Manchester_2008, Toeroek_2018, Verbeke_2019}, they are too computationally expensive and unwieldy for analyzing the magnetic field measurements of CME events}. For this purpose multiple semi-empirical models have been developed \citep{Isavnin_2016, Moestl_2018, kay_2018, rouillard_2020, Weiss_2021}, that aim to mimic the general properties that ICMEs are thought to have. These models serve as a bridge in between the more complex MHD models and the fully rigid analytical models and can include simple interaction with the solar wind, some form of deformations, and an internal magnetic field while retaining some of the computational simplicity of the simpler analytical models.

In our previous paper \citep{Weiss_2021}, henceforth referred to as W21, we showcased the successful application of the semi-empirical 3D coronal rope ejection (3DCORE) model to a single point event detected by Parker Solar Probe. The 3DCORE model is a semi-empirical forward simulation model that describes an ICME as a self-similarly expanding torus-like structure that remains attached to the Sun. It includes ICME evolution factors such as flux rope expansion and deceleration due to slower solar wind which are described using empirical relations. The internal magnetic field structure is based on an analytical toroidal uniform-twist solution described in \cite{Vandas_2017}.

In this paper we attempt the novel approach of simultaneously fitting multiple magnetic field measurements from different spacecraft at different locations. So far the majority of multi point studies have only compared different single point fits from different spacecraft from the same underlying event \citep{Moestl_2012}. The results from our analysis will show if it is even possible to describe multiple observations using the global assumptions within our model and, if not, give a hint on its deficiencies. \textbf{The comparison of the single point fits and multi point fits of the same event also allows for a sort of cross-validation which can further the understanding on the underlying event and the used model}.

\textbf{Section \ref{sec:event} introduces the two events for which we will test our approach}. The first event, the April 19 ICME, represents three very close by high-quality measurements for which we would expect our approach to have the highest chances of success. The second event, the May 28 ICME, presents a bigger challenge and was primarily chosen to test the limits of our methods. In Section \ref{sec:model} we give a brief recap of the 3DCORE model from W21 and also describe the two changes that were made to update the model. Section \ref{sec:methods} describes the adapted ABC-SMC fitting algorithm that was used for fitting the multi point data. Section \ref{sec:results} shows the reconstructed magnetic field measurements for both events and the inferred model parameters from multiple spacecraft combinations. Finally, in Section \ref{sec:conclusion} we discuss the results and the conclusions we can draw.

\section{Data and Events} \label{sec:event}

\begin{figure*}[ht]
\center
\includegraphics[trim=50px 50px 20px 50px, clip,width=.48\linewidth]{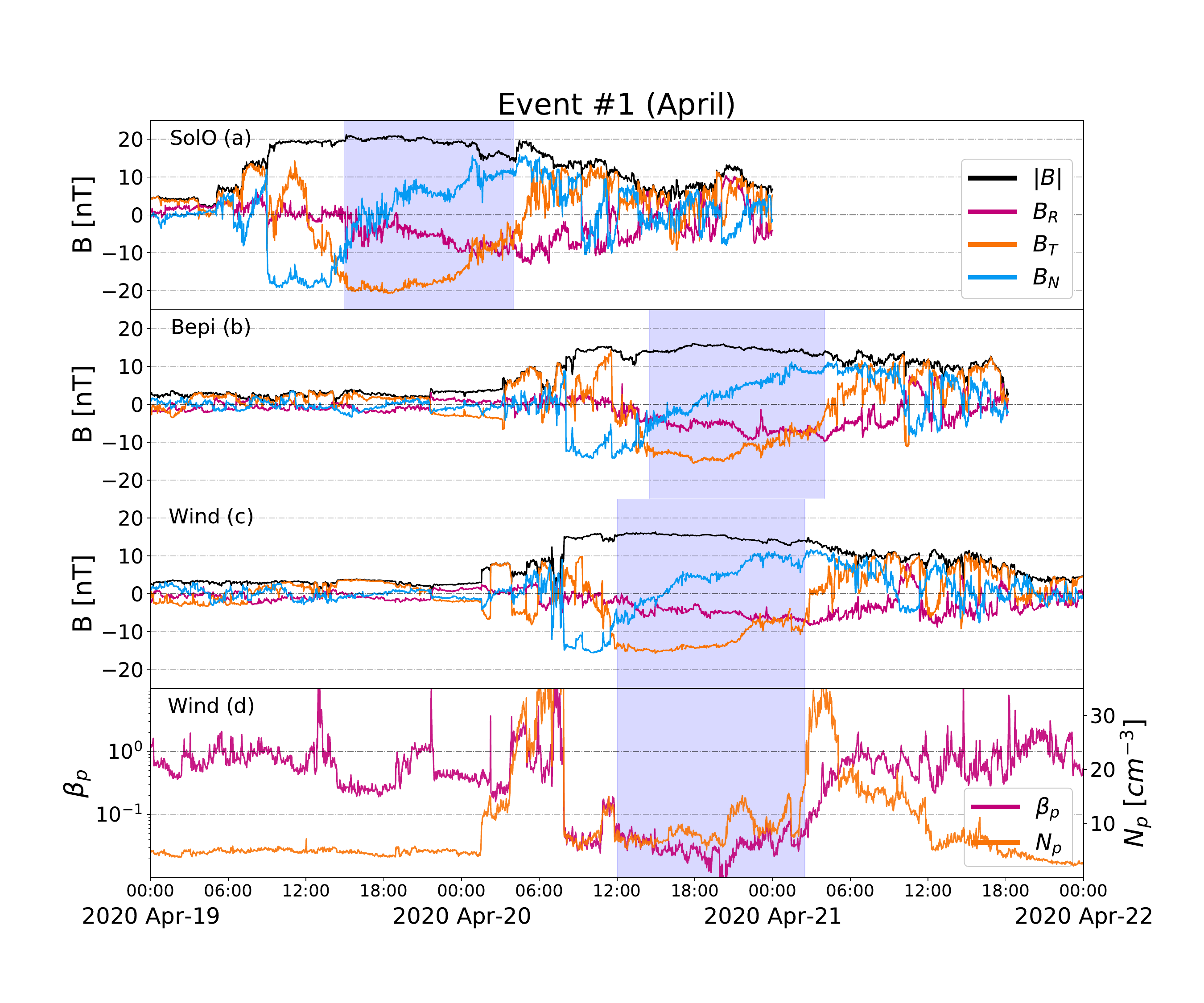}
\includegraphics[trim=50px 50px 20px 50px, clip,width=.48\linewidth]{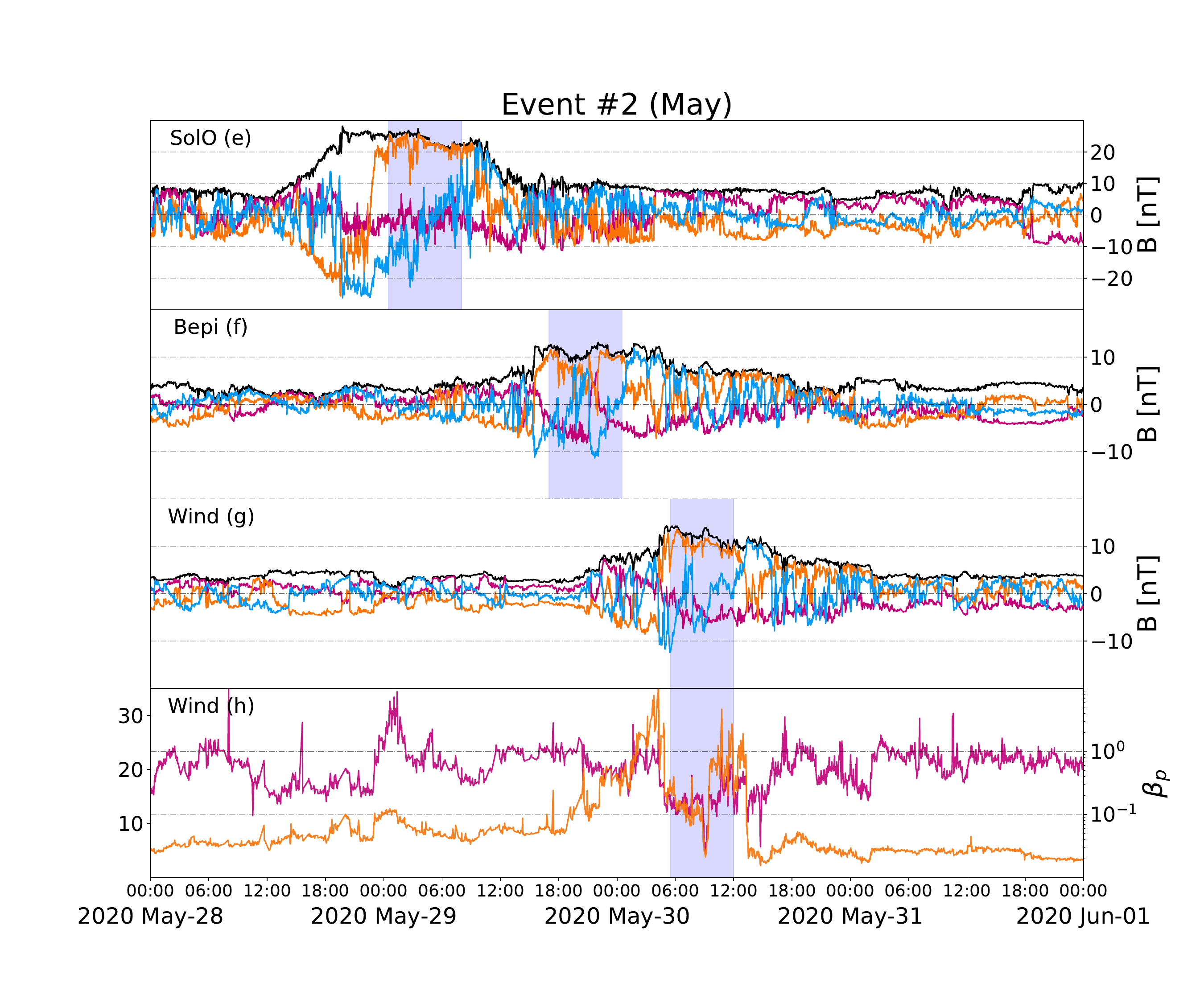}
\caption{\label{fig:events_insitu} In situ magnetic field measurements in local radial tangential normal (RTN) coordinates from Solar Orbiter (\ref{fig:events_insitu}a/e), BepiColombo (\ref{fig:events_insitu}b/f) and Wind (\ref{fig:events_insitu}c/g) for the 2020 April 20 (left) and 2020 May 28 (right) event. The two bottom panels (\ref{fig:events_insitu}d/h) show the in situ proton density number $N_p$ and the associated plasma beta coefficient $\beta_p$. The blue shaded area represents our estimated intervals for the unperturbed magnetic flux ropes (UMFR) which we will be subsequently using for our fitting approach.}
\end{figure*}

To demonstrate the applicability of our 3DCORE model and the associated Monte-Carlo based fitting method for modeling the magnetic field of multi point ICMEs, we choose two recent events that were both observed in situ by the Solar Orbiter (SolO), Wind and BepiColombo (Bepi) spacecraft.

The first and primary event that we will make use of is the 2020 April 19 ICME which we henceforth refer to as the April event. This event was initially detected in situ by Solar Orbiter at a heliocentric distance of  $0.80$~AU on 2020 April 19 05:06 UTC, at about 4 degrees longitude (HEEQ) east of Earth. The ICME shock subsequently arrived a day later at Wind and BepiColombo on the April 20 at 01:34 UTC and 03:09 UTC respectively. For an extensive description of this event we refer the reader to \cite{Davies_2021} as we will only focus on the details that are of interest for our analysis.

The second event that we have chosen is the 2020 May 28 ICME which we henceforth refer to as the May event. Similar to the April event, this ICME was again first detected by Solar Orbiter at a significantly smaller heliocentric distance of $0.55$~AU on May 28 15:00 UTC. The ICME was again also observed by Wind and BepiColombo at a very large longitudinal separation on May 29 at 15:15 UTC and May 30 at 01:15 UTC respectively. Due to the lack of clear indicators of a shock front these arrival times represent rough estimates.

\textbf{Figure \ref{fig:events_insitu}} shows the in situ measurements from the SolO MAG magnetometer \citep{Horbury_2020}, the Wind MFI magnetometer \citep{Lepping_1995}, the Bepi MPO-MAG magnetometer \citep{Heyner_2021} and the Wind SWE plasma instrument \citep{Ogilvie_1995} for both events. The magnetic field measurements are shown in their respective local radial-tangential-normal (RTN) coordinate systems. For the plasma measurements from the Wind spacecraft we only show the proton density $N_p$ and the associated computed plasma beta coefficient $\beta_p$. The blue shaded area represents our estimated intervals for the unperturbed magnetic flux ropes (UMFR, see \cite{Davies_2021} for the April event) which will be the intervals that we use for our fitting analysis. The UMFR intervals were estimated using both the magnetic field measurements and the plasma measurements and matched with each other using features in the magnetic field data.

\begin{figure*}[h]
\center
\includegraphics[trim=175px 125px 125px 185px, clip,width=.48\linewidth]{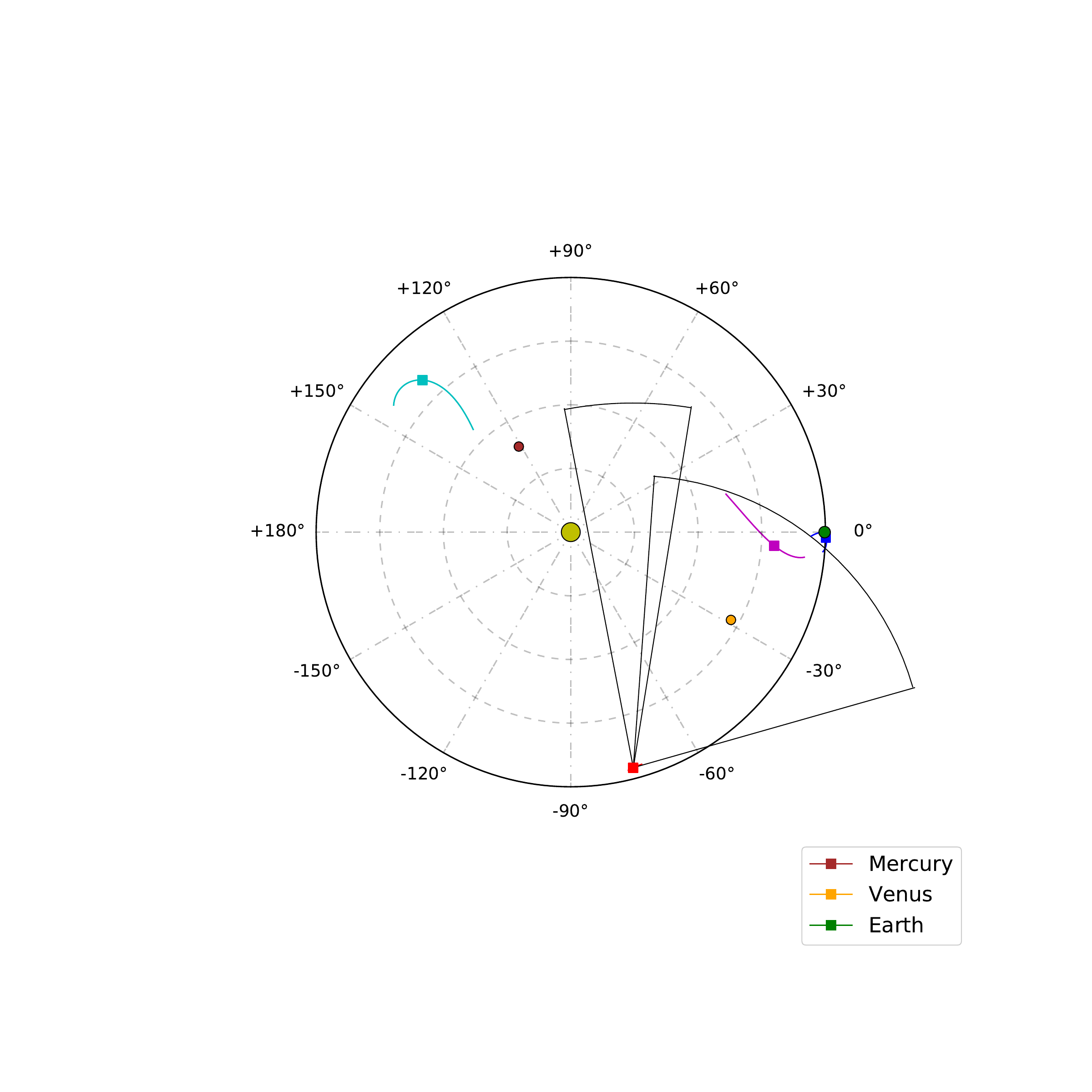}
\includegraphics[trim=175px 125px 125px 185px, clip,width=.48\linewidth]{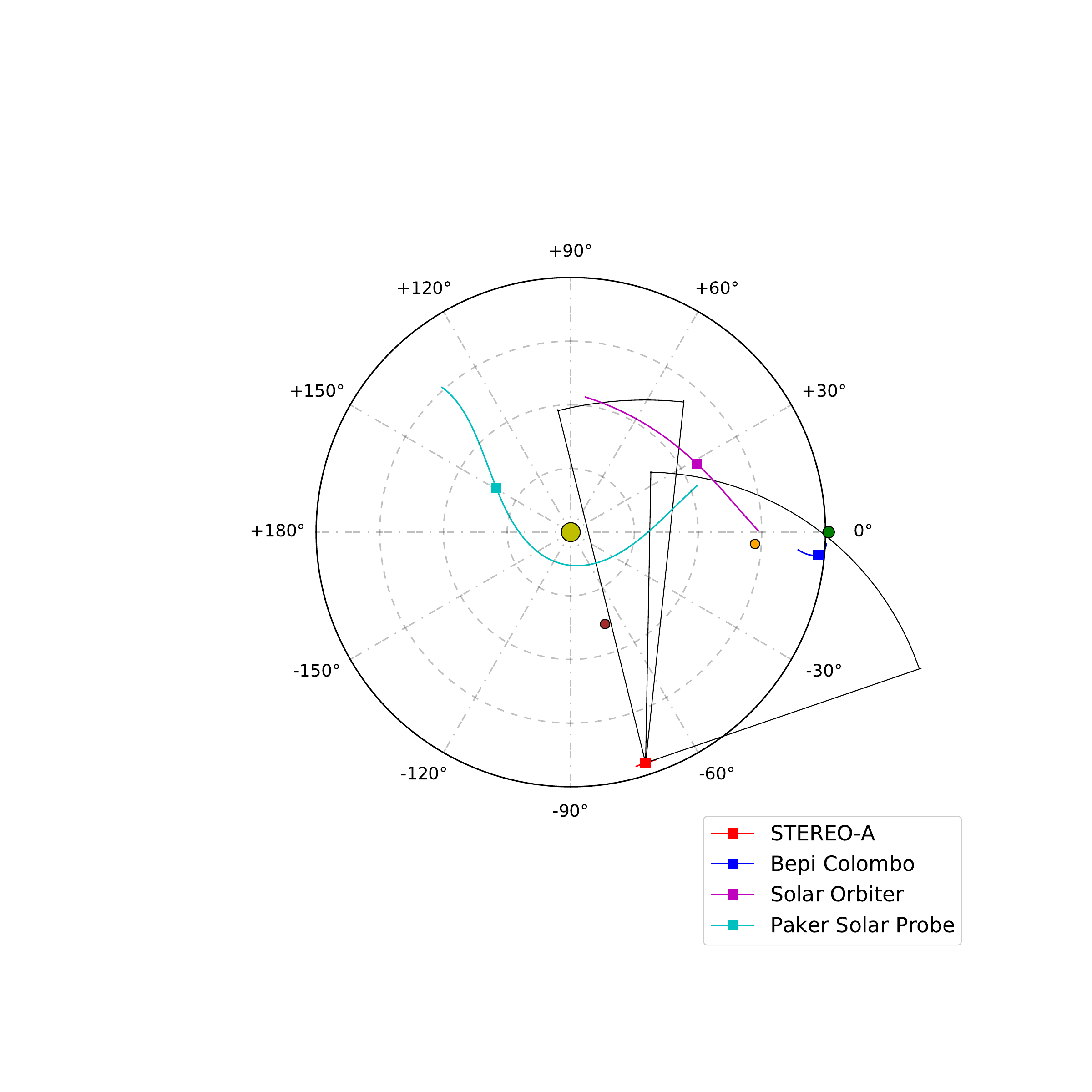}
\caption{\label{fig:events_overview} Spacecraft positions for Solar Orbiter, BepiColombo, Parker Solar Probe and STEREO-A on 2020 April 20 00:00 UTC (left) and on 2020 May 29 00:00 UTC (right) in HEEQ coordinates. Furthermore shown are the planets Mercury (brown), Venus (orange) and Earth (green). Colored trails show orbital movements of the spacecraft within a time range of two months around the specifically chosen position. Two open ended cones show the field of view of the HI-1 ($20^\circ$) and HI-2 ($70^\circ$) imaging instruments aboard STEREO-A.}
\end{figure*}

\textbf{Figure \ref{fig:events_overview}} shows the different spacecraft positions within the heliosphere on 2020 April 20 00:00 UTC and 2020 May 29 00:00 UTC including other satellites such as Parker Solar Probe (PSP) and the inner planets. As can be seen from these two figures the STEREO-A spacecraft, which carries heliospheric imagers \citep[HI,][]{Eyles_2009}, was ideally positioned at both times to image and track Earth-directed CMEs. The captured images for both events are briefly discussed in Section \ref{sec:event_remote}. Unfortunately, PSP was located on the back side of the Sun and therefore incapable of recording any of these two events. \textbf{Table \ref{tab:events_position}} summarizes the positions of all three spacecraft that observed the events in situ on the same dates in Heliocentric Earth equatorial (HEEQ) coordinates.

\begin{table}[h]
\caption{Positions of Solar Orbiter, Wind and BepiColombo on the 2020 April 20 00:00 UTC (top) and 2020 May 29 00:00 UTC (bottom) in HEEQ coordinates.}
\label{tab:events_position}
\centering                  
\begin{tabular}{c r | c c c}    
& & Solar Orbiter  &  Wind  & BepiColombo \\ 
\hline\hline
\multirow{3}{*}{\rotatebox[origin=c]{90}{April}} & r $\hphantom{d}$[$au$] & $0.80$ & $0.99$ & $1.01$\\
&lon [$deg$] & $-3.80$ & $0.18$ & $-1.23$\\
&lat [$deg$] & $-3.85$ & $-5.15$ & $-5.54$\\
\hline
\multirow{3}{*}{\rotatebox[origin=c]{90}{May}} & r $\hphantom{d}$[$au$] & $0.55$ & $1.00$ & $0.97$\\
&lon [$deg$] & $30.11$ & $-0.14$ & $-5.26$\\
&lat [$deg$] & $4.70$ & $-0.91$ & $-2.70$\\
\end{tabular}
\end{table}

The magnetic obstacle of the April event appears as an almost pristine flux rope in all three in situ measurements with a nicely rotating magnetic field. The normal component $B_N$ is bipolar, the transversal component $B_T$ unipolar and the radial component $B_R$ negative near zero. According to the standard classification schemes this corresponds to a low inclination left-handed South-East-North flux rope \citep{Bothmer_1998, Davies_2021}. Due to the textbook like structure of this flux rope it represents a unique opportunity to test the validity of our model with multiple close-by in situ measurements. 
The May event shows significantly more complex structure and was primarily chosen to identify the limits of our modeling approach at extremely large longitudinal separations. While one can still clearly identify a right-handed South-West-North (SWN) flux rope in the SolO MAG measurements, which represents a low inclination flux rope with the opposite axial direction and handedness compared to the April event, the magnetic field is strongly perturbed at the center for Bepi and Wind. Nonetheless, taking only the endpoints of the Bepi and Wind measurement into account the magnetic field still exhibits a magnetic field rotation that is comparable to Solar Orbiter. The differences can be explained due to large spatial distance in between the spacecraft and poses a significant challenge for our modeling efforts. \textbf{We will partially alleviate these issues by applying strong smoothing onto the magnetic field data from the May event to suppress the smaller scale perturbations and only using the less perturbed endpoints for our analysis.}

\subsection{Remote Observations} \label{sec:event_remote}
 
 \begin{figure}[h]
\center
\includegraphics[trim=150px 50px 150px 110px,  clip,width=\linewidth]{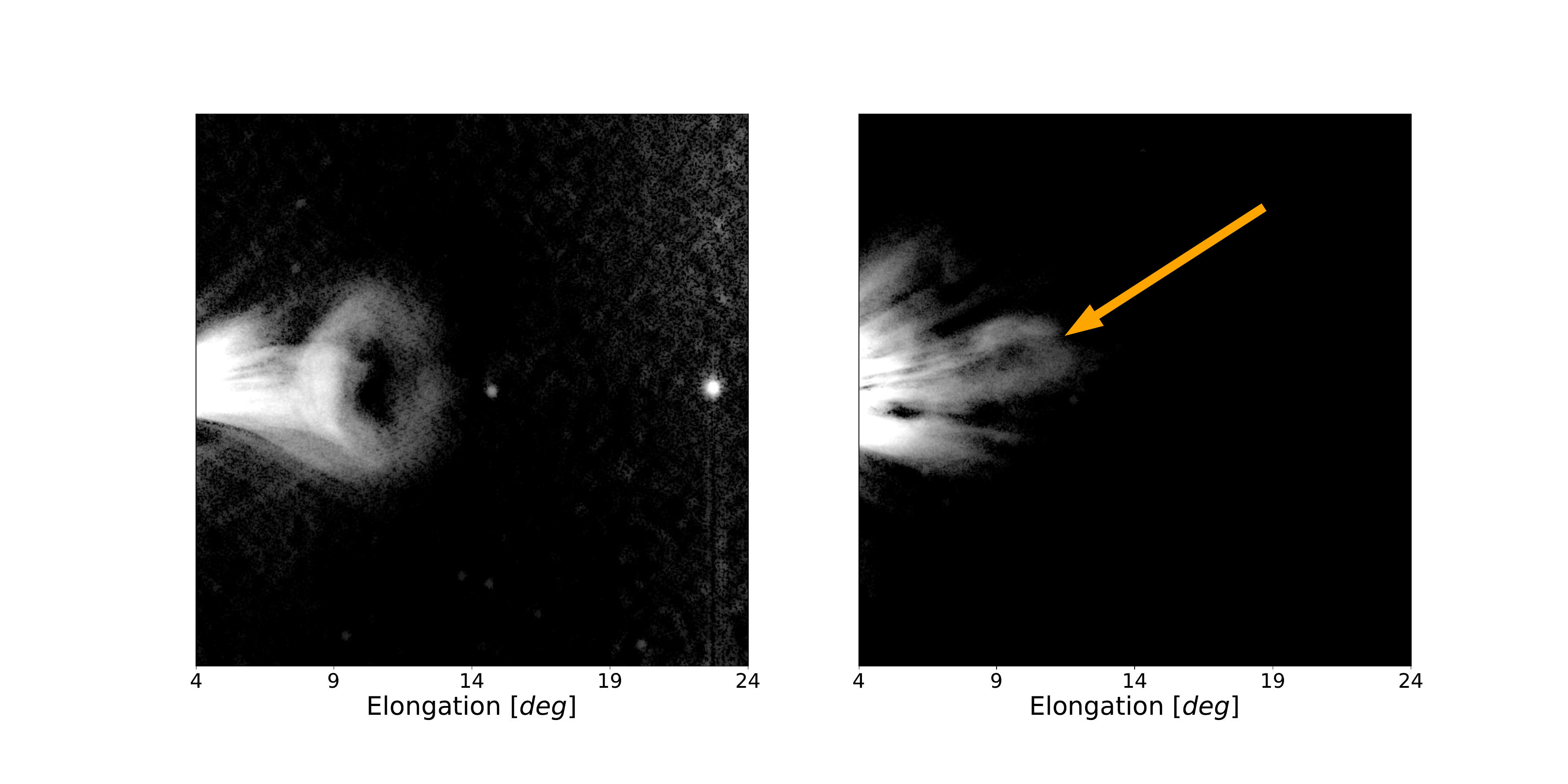}
\caption{\label{fig:events_hi} White light images, with the stars subtracted using a median filter, from the STEREO-A HI-1 instrument showing the April event on the 16 of April 07:29 UTC (left) and the May event on the 26 of May 15:29 UTC (right). The image of the April event shows a slightly flattened Earth-directed CME with a faint leading edge, elliptical cavity and a strong density pile up at the back. The image of the May event shows a structure at a higher latitude that is hard to identify in a single image (marked with the yellow arrow) and does not contain an obvious cavity.}
\end{figure}

\textbf{Figure \ref{fig:events_hi}} shows two white light images taken by the HI-1 instrument on 2020 April 16 07:29 UTC (left) and 2020 May 26 15:29 UTC (right). The HI-1 instrument was able to nicely follow the evolution of the April CME over a time range of almost 24 hours showing a faint leading edge, an elliptical cavity and a strong density pile up at the trailing edge. The CME cavity, which represents the magnetic obstacle of interest, is seen edge-on and appears to be of elliptical shape with a time-invariant aspect ratio of around 2 \citep[see][]{Davies_2021}. These images for the April event appear to be consistent with the approximation of an elliptical cross-section that is one of the central assumptions built into our model. The white light images for the May event on the other hand show a smaller and fainter structure propagating above the ecliptic. The structure also darkens considerably faster making it impossible to observe the evolution of the shape for a longer duration than a few hours.

Both the April\footnote{\url{https://www.helcats-fp7.eu/catalogues/event_page.html?id=HCME_A__20200415_01}} and May\footnote{\url{https://www.helcats-fp7.eu/catalogues/event_page.html?id=HCME_A__20200526_01}} event are part of the HELCATS catalog. Despite only being observed from a single vantage point one can extract rough estimates for the propagation speed and direction by tracking the CME evolution. Under the assumption of self-similar expansion the given values for the April event are an average CME speed of $339\pm11\,\textrm{kms}^{-1}$ and a propagation direction of $-2\deg$ longitude and $-6\deg$ latitude in HEEQ coordinates. For the May event one receives an average speed of $497\pm64\,\textrm{kms}^{-1}$ and a propagation direction of $36\deg$ longitude and $11\deg$ latitude (HEEQ).

\section{Model} \label{sec:model}

In this paper we use an updated version of the 3DCORE model to analyze our two selected events. While we will give a brief general overview we will focus on the changes to the implementation that we have made in order to facilitate analyzing multi point events. The two primary changes that are worth mentioning are the ``twist reparametrzitation'' where we substituted the magnetic field twist parameter using a different related parameter which has beneficial numerical properties and the introduction of a new noise model. For the interested reader we refer to W21 for a detailed description of the original implementation.

\subsection{3DCORE Overview}

The 3DCORE model is an empirically based forward simulation model which can generate synthetic in situ magnetic field measurements anywhere within the heliosphere. It assumes a self-similarly expanding torus-like MFR structure that propagates along a fixed direction radially away from the Sun. The end points of the torus-like structure stay attached to the Sun. The expansion of the flux rope structure is mimicked using a simple scaling relation. The interaction with the background solar wind is described by using a drag-based model \citep{Vrsnak_2013} where the frontal part of the structure is accelerated or decelerated, depending on the relative speed to the ambient solar wind speed. The solar wind speed is assumed to be constant everywhere. The ``pancaking'' effect, that is described in \cite{Riley_2004A}, is approximated by stretching the cross-section of the torus into an elliptical shape.

\begin{figure}[h]
\center
\includegraphics[trim=325px 375px 275px 375px, clip,width=\linewidth]{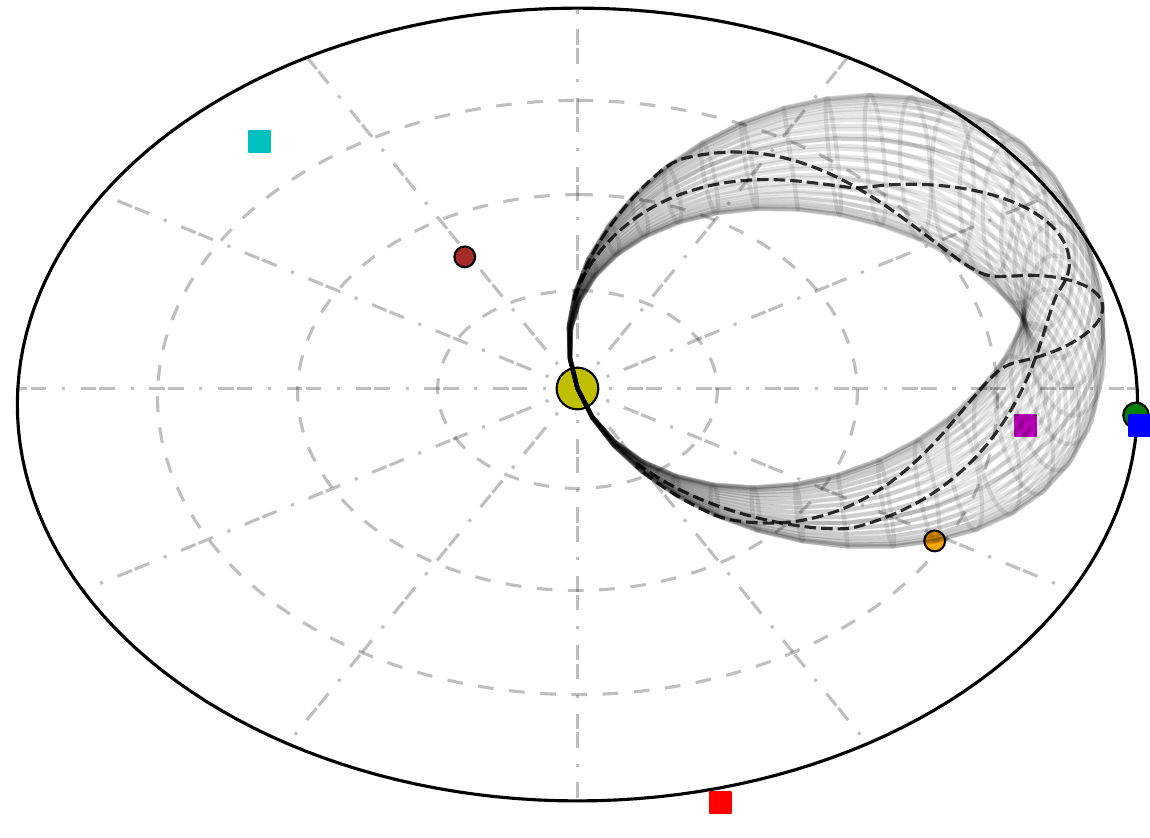}
\caption{\label{fig:3dcore_model} Example of a 3DCORE simulation showing the torus-like structure at near $1$~AU with a highly elliptical cross-section and two magnetic field lines at differing radial distances (twist number $\tau = 5$).}
\end{figure}

The internal magnetic field of the MFR is simulated by inserting an analytical magnetic field model into the structure. We use a model based on the solutions described in \citep{Vandas_2017} which we have slightly modified in order to accommodate the elliptical cross-section. \textbf{We furthermore make the assumption that the total number of twists along the torus stays constant in time, which leads to a decreasing twist per unit length as the structure expands.} The decrease of the magnetic field strength as the structure propagates forward and expands is again described by a simple scaling relation \citep{Leitner_2007}. While the implemented magnetic field is based on analytically derived equations, it is important to note that it is not fully physical and violates various conservation laws. We assume that the errors that are introduced by this approximation are less severe than those arising from the basic assumptions on the geometry.

One of the key points of our model is that the simulation is extremely fast as it is only based on empirical relations and does not solve any expensive and complicated MHD equations. This efficiency makes it significantly easier to explore the full parameter space of the model and allows the usage Monte-Carlo based methods which require in the order of thousands or millions of simulation runs.

In total the model contains 12 controllable parameters: the initialization time $t_0$, propagation direction and orientation $lat,\ lon,\ inc$, cross-section width $d_{1au}$, cross-section aspect ratio $\delta$, initialization distance $r_0$, initial speed $v_0$, magnetic field strength $b_{1au}$ at 1AU, twist number $\tau$, global solar wind speed $v_{sw}$ and drag parameter $\Gamma$. Both the diameter of the cross-section and magnetic field strength are modified by scaling relations with $d(t) = d_{1au} r(t)^{1.14}$ and $b(t) = b_{1au} r(t)^{-1.68}$, where $r(t)$ is the distance of the CME front from the Sun.

\subsection{Twist Reparameterization}

One of the key findings of W21 was that we were not able to accurately infer the magnetic field twist $\tau$ and the aspect ratio of the torus cross-section $\delta$ using single-point observations. We found a strong $\tau - \delta$ degeneracy meaning that these two parameters are strongly correlated. The dependency relation roughly follows $\tau \propto \delta / E(\delta)$ where $E(\delta)$ is the function for the circumference of an ellipse with a given aspect ratio.

This issue exists as we are not directly capable of measuring the twist of the magnetic field, and instead only measure the local direction of the magnetic field vectors. Depending on the geometry or size, in this case the height of the CME structure given by the aspect-ratio, the total number of twists along a fixed length can change drastically. On top of the issue of not being able to accurately infer these model parameters simultaneously we found this degeneracy to be a numerical issue when using our Monte-Carlo based fitting algorithm as the degeneracy is strongly non-linear and is inefficiently sampled using multivariate Gaussian transition kernels. The shape of this degeneracy, as seen in a two dimensional histogram, closely resembles a ``banana shape'' and presents a classical problem for the class of algorithms that we use \citep{Haario_1999, Filippi_2011}.

We therefore attempt to at least remove the non-linearity of this degeneracy by defining a new twist parameter $T_\tau = \tau E(\delta)$. The introduction of this new parameter significantly reduces the numerical instability of our approach and increases the overall speed of the fitting algorithm as the sampling efficiency is increased. The original $\tau$ values remain recoverable in a post-processing step.

\subsection{Noise Model}\label{sec:noise_model}

In order to estimate the uncertainties within our model with our fitting algorithm we need to be capable of simulating the measured noise. In the context of our model we gather all unwanted small scale components of the magnetic field under the general term of ``noise''. This includes the expected instrument noise but also other seemingly random phenomena such as MHD turbulence and waves or fluctuations due to undetermined physical processes. In the case of instrument noise, we assume that the data sets are thoroughly cleaned and that the most significant spacecraft influences are correctly removed. 

In W21 we made the assumption of a simple additive Gaussian noise model that was described by a single parameter $\sigma$ which determined the standard deviation of the additive noise component. As was briefly mentioned in W21, this is a very basic approach which is easy to implement but fails to fully capture the small scale behaviour of the magnetic field that is seen in the observations. We will therefore introduce a more sophisticated additive noise model based on the power spectrum distribution (PSD) $P(k)$. Our implementation for generating the noise is inspired by the analogous process that is used to generate primordial perturbations for dark matter \textit{N}-body simulations using the matter power spectrum \citep[see][]{Hahn_2011}.

For our purpose, we are interested in using the PSD to reproduce the type of fluctuations within our model that are seen in the actual observations. As a first step we require an appropriate PSD from which we generate the noise. The most adequate choice is the PSD from the underlying event which is being analysed as the distributions can vary significantly in between different events. In our case we ignore cross-correlations between the three different components of the magnetic field which significantly simplifies the process as we only need to generate three independent noise samples for each component. Given a random noise field $\mu(t) \sim \mathcal{N}(0, 1)$ and its corresponding Fourier transform $\Tilde{\mu}(k) = \mathcal{F}\left(\mu(t)\right)$ one can generate a random time series:
\begin{equation}
\delta(t) = \mathcal{F}^{-1}\left(\sqrt{P(k)} \Tilde{\mu}(k)\right)
\end{equation}
so that $\Tilde{\delta}(t) \approx P(k)$. This process allows us to generate random and unique noise fields which all have the same underlying PSD statistic. In the final step we add this random field to our model output to combine the large scale magnetic field rotation with the random small scale fluctuations.

We compute the PSD for each magnetic field component independently using Welch's algorithm \citep{Welch_1967} from the time interval of interest. In order to remove the large scale fluctuations that partially describe the magnetic field rotation that is described by our flux rope model we pre-process the measurements using a linear de-trending process on an hourly scale.

As the PSD varies only little for each component we generate a single general distribution from the average over all three components. In the next step we use the computed PSD to generate a random sequence with the same underlying PSD that will act as our noise. As white noise (Gaussian noise) has a flat power spectrum we can generate the desired sequence by first generating a random Gaussian sequence $\delta \sim \mathcal{N}(0, 1)$ and multiplying the newly generated sequence in Fourier space with $P(k)$ before transforming the sequence back into real space. This process is performed independently for each component of the magnetic field on top of which the noisy $\delta$ sequence is then added.

The biggest advantage of this newer noise model is that it eliminates the necessity of a noise level parameter as the severity of noise is directly estimated from the measurements. In theory this approach also allows us to fit profiles with an extremely high density of fitting points as the the point-to-point fluctuations are more accurately modelled to decrease with increasing proximity. In practice this is not done as the inaccuracies are dominated by the assumption of the geometry and magnetic field model.

\textbf{While this noise model is expected to be a large improvement over the model employed in W21 it is still important to note one of its deficiencies. The fluctuations in the magnetic field strength, which represent compressive fluctuations, are generally weaker than those of its components that describe 
incompressible fluctuations \citep[e.g.][]{Good_2020}. This is a strong hint that the underlying fluctuations would be better described using rotations, rather than the additions that are used in our noise model. But as our summary statistic that we use for our fitting algorithm only makes use of the magnetic field components and not the total field we expect the impact to be negligible.}

\section{Multi point fitting algorithm} \label{sec:methods}

In this section we focus on the details of the fitting algorithm that we use to analyze the magnetic field measurements with our 3DCORE model. The underlying algorithm, an approximate Bayesian Computation sequential Monte-Carlo (ABC-SMC) fitting algorithm, remains largely unchanged from W21 and requires only minimal adaptions for the multi point scenario. 

The general idea behind the ABC-SMC fitting algorithm is to generate large ensemble simulations from an initial parameter space and classify the goodness of fit for each simulation run according to an error metric, called the summary statistic, and reject all simulations which fall above a certain error threshold. An important step is also the rejection of all simulations that do not generate valid results at all due to the underlying CME geometry not hitting any of the observers or producing magnetic field signatures that are significantly time shifted due to the simulated CME arriving too late or early.

By repeating this procedure and using the accepted simulations as a representation of the initial parameter space for the next iteration one can sequentially reduce the error threshold, starting from a larger value, and arrive at a good estimate of the underlying posterior after a certain number of iterations.

Due to the stochastic nature of the simulations, as we include the noise model introduced in Section \ref{sec:noise_model}, and the correct implementation of an ABC algorithm we are able to interpret the generate ensembles as probability distributions (in the Bayesian view). This naturally gives us not only best-fit model parameters but also non-Gaussian error estimates and parameter correlations which can further deepen the understanding of the results and the underlying model.

We discuss the changes that we have made to the summary statistic and the pre-filtering process for determining if a simulation run can be considered as valid or invalid. We also take a brief look at the inherent biases that may appear in our analysis.

\subsection{Summary Statistic}

For each single observation we use a normalized root mean square error statistic between model output $x$ and reference observation $y$ which is defined as followed:
\begin{equation}
\rho(x, y) = \sqrt{\frac{\sum_{i=1}^K (x(t_i) - y(t_i))^2}{\sum_{i=1}^K y(t_i)^2},  }
\end{equation}
where $t_i$ is one of the $K$ time points at which the magnetic field is to be compared. 

At this point it may appear viable to create a combined statistic that condenses multiple statistics into a single number such as the mean RMSE over multiple observations. In practice we have found that this approach runs into the issue in which the algorithm can make a hidden trade-off in between improving the fit for one observation at the expense of the others. For this reason we adapted our method so that it simultaneously optimizes multiple error metrics. While this does not necessarily guarantee optimal convergence it does not allow for a scenario where one observation is optimized at the negative trade-off of another and the end result clearly shows if a certain observation cannot be properly fitted.

\textbf{As a consequence, a simulation output for N observations $x = (x^1, \hdots, x^N)$ is accepted if and only if $\rho(x^k, y^k) < \epsilon^k$ for all $k \in [1, N]$. The iterative threshold values $e^k$ are computed in the same way as in W21 with $e^k_0 = \rho(0, y^k)$ and $e^k_{j+1}$ is set as a quantile fraction of $e^k_{j}$.}

\subsection{Adaptive Time Markers} \label{sec:bias_degen}

An important aspect of the fitting algorithm is to determine if a simulation run can be considered as valid in the first place. In a large ensemble simulation a majority of runs will not generate any valid results due to missing one or many of the defined observers or producing signatures that are strongly time-shifted when compared with the reference observation that is to be reproduced. In W21 we used two time markers, $t_s$ and $t_e$, to determine valid simulation runs.

The condition that we initially set was that any simulation run is to be considered valid if and only if it produces a valid magnetic field signature at any of the time points $t_i$ that are part of the summary statistic and produces an invalid measurement at $t_s$ and $t_e$. This requires the observer to be within the CME geometry, as assumed by the 3DCORE model, at any time point $t_i$ and to be outside of it at $t_{s/e}$. By setting $t_s$ just before $t_1$ and $t_e$ just after $t_K$ one is able to closely control in situ duration of any single valid run.

The problem that arises, specifically in the multi point scenario, is that it is initially very hard to find simulations which produce signatures of just the right length. As one adds additional constraints by adding observations the efficiency of the fitting algorithm can suffer significantly. For this reason we implemented an adaptive time marker scheme, in which the the $t_s$ and $t_e$ markers are set at a large temporal distance to the the fitting time points $t_i$. This distance is iteratively reduced, as is the error threshold, until it reaches a final value after a certain number of iterations. This considerably helps the sampling process in the initial stages when the explored parameter space is still large. From basic tests we have concluded that this initial approximation does not have any significant negative effects on the final results.

\subsection{Bias} \label{sec:bias_corr}

An inherent property of our 3DCORE model is that it is an empirical model and does not contain any physical calculations. As we will show, this can lead to some peculiar results which require some additional treatment. While our fitting algorithm and error metric guarantees that we only select model runs and their associated model parameters that match the observations well it can only make use of ``valid'' runs. Simulation runs that do not generate any results due to the 3DCORE shape not hitting any of the predefined observers or runs that generate magnetic fields that are too strongly time shifted are automatically rejected before computing any error metrics. 

While the size of the shapes is limited by using bounds for the model parameters this nonetheless leads to an extremely strong bias towards the largest allowed shapes as they do not fall victim to this selection bias. The two parameters in our model that control the size of our structure are the $d_\text{1AU}$ (torus width) and $\delta$ (torus height) parameters. The width parameter $d_\text{1AU}$ is constrained by the duration of the in situ measurement and can therefore be more or less estimated by our algorithm but the height is completely unknown due to all observers being largely located within the ecliptic plane. Due to the $\delta-\tau$ degeneracy that we mention previously we know that we cannot constrain the $\delta$ parameter using a single in situ measurement. This generally leads to linear bias $p(\delta) \propto \delta$ as a structure that has a larger aspect ratio $\delta$ has a larger chance of hitting the observer.

\textbf{Under certain circumstances this bias does not occur, such as a CME that directly propagates towards the observer, which makes this issue hard to handle.} In general this must be done at a case by case basis by looking at the model simulations that are accepted and rejected due to hitting or missing the observers. The conclusion is that our analysis will normally generate extremely large estimates for the $\delta$ parameter while the correct interpretation is that nothing about this parameter is actually known. One hope is that this bias would disappear when adding multiple observations which would largely resolve this issue.

\section{Results} \label{sec:results}

\begin{figure*}[h]
\includegraphics[trim=35px 75px 100px 65px, clip,width=.48\linewidth]{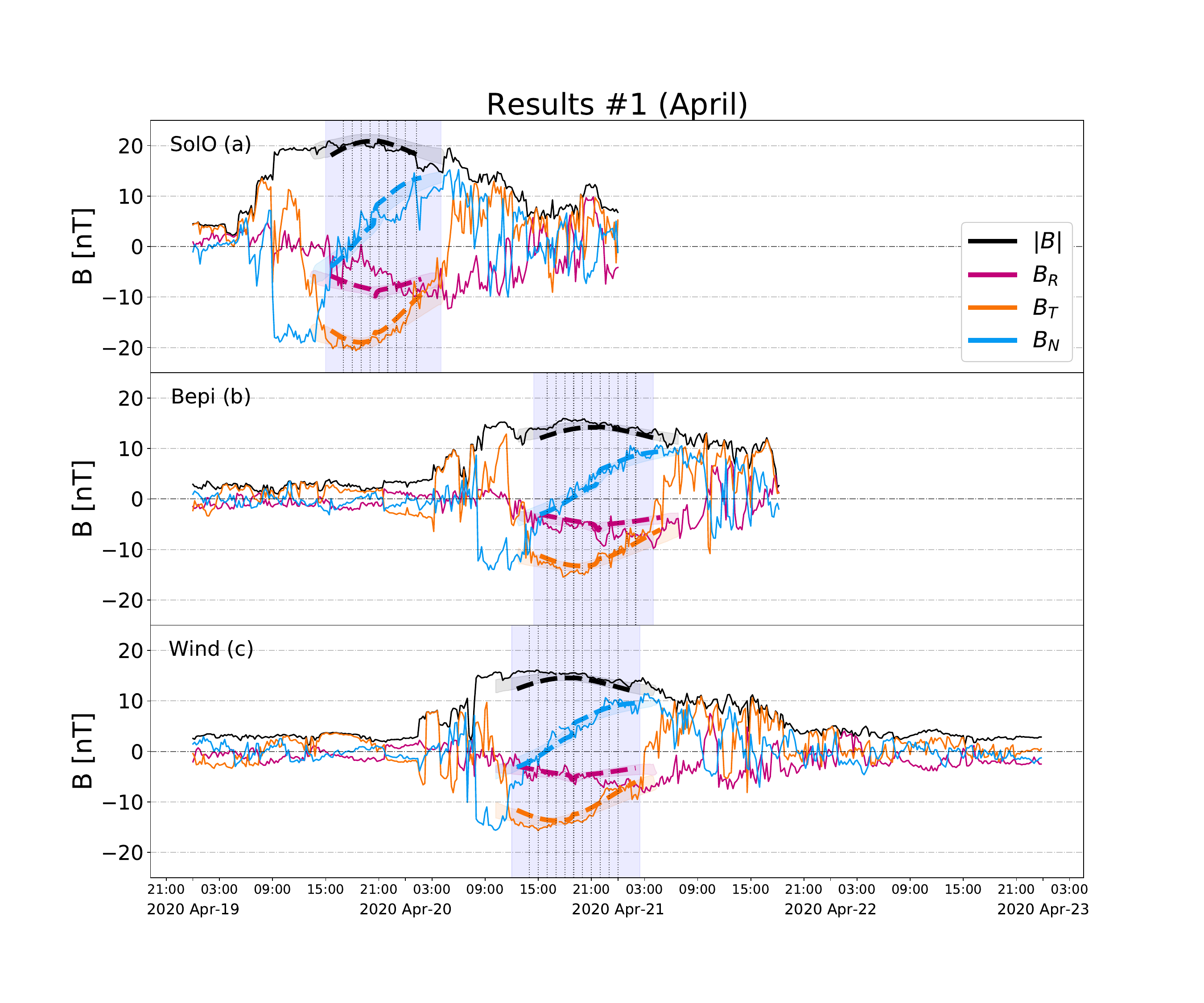}
\includegraphics[trim=100px 75px 35px 65px, clip,width=.48\linewidth]{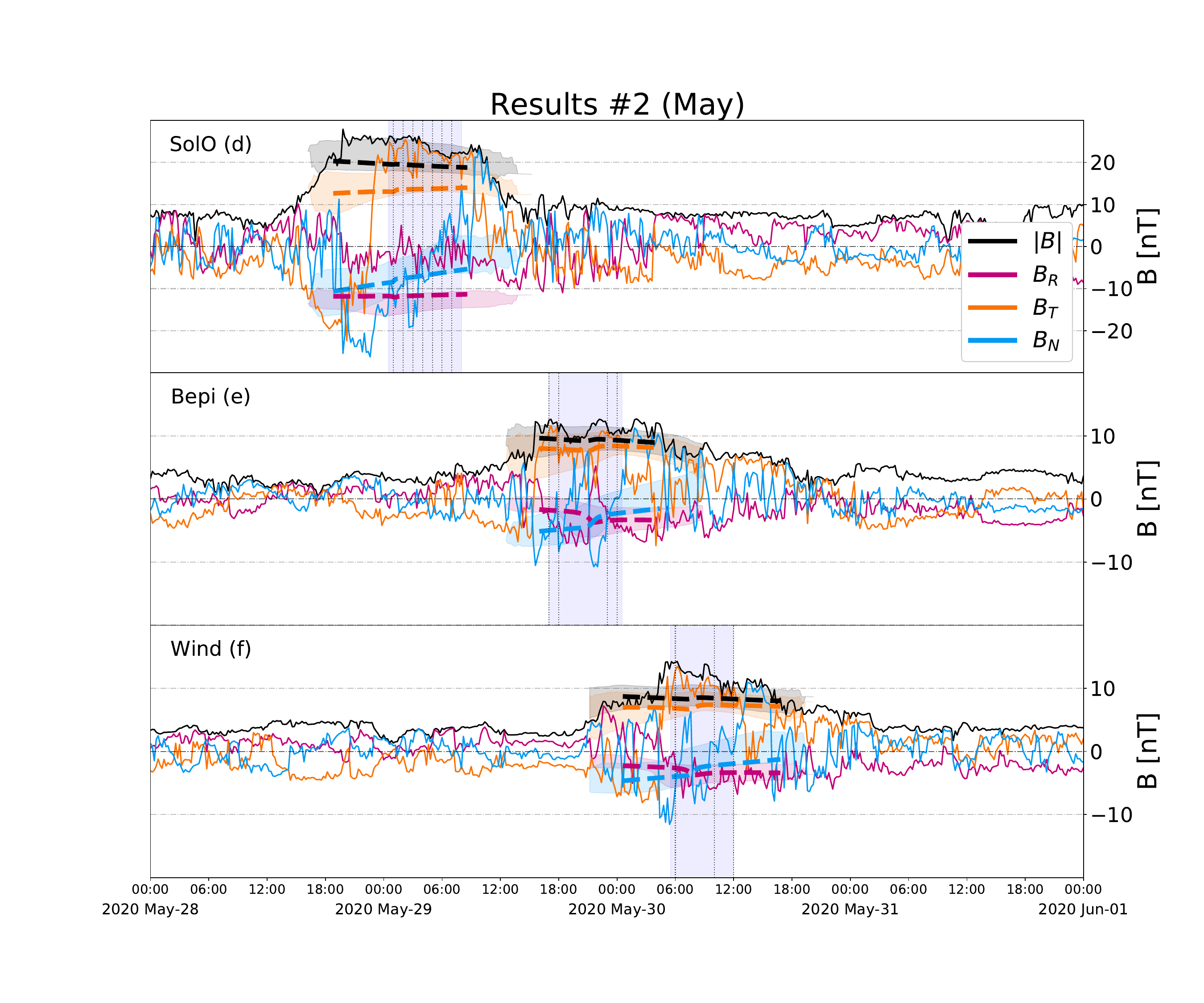}
\caption{\label{fig:results_insitu} Reconstructed synthetic magnetic field measurements (dashed) with $1\,\sigma$ confidence interval (shaded area) using the triple point constellation for the April event and a dual-point constellation for the May event. The reconstructed magnetic field is overlayed on top of the real in situ measurements for Solar Orbiter (a/d), BepiColombo (b/e) and Wind (c/f). The blue shaded area represents the defined UMFR interval and the dotted vertical lines show the used fitting points $t_i$. All magnetic fields are shown in local RTN coordinates.}
\end{figure*}

In this section we present the results of our paper in which we fitted the April and May event using our updated 3DCORE model and ABC-SMC fitting algorithm. A constellation of the three different satellites, combined with their in situ measurements, not only allows us to create a single large analysis encompassing all three satellites but also three dual-point analyses in which a combination of two of the three available in-situ measurements are used. Counting the single-point results this then leads to a total of seven different possible data combinations which can be used. This allows us to study in detail how the results using our method differs when using either the full data set or smaller sub sets.
   
\textbf{Figure \ref{fig:results_insitu}} shows the resulting triple-point fit of the 2020 April 19 CME using all three in situ measurements and a dual-point fit of the 2020 May 28 CME for SolO and Wind in local RTN coordinates. While we were capable of producing a triple-point fit for the April event we did not achieve satisfactory results for the May event. The dashed lines show one single representative fit from the generated ensemble. The shaded area around this single fit shows the 1-$\sigma$ spread of all ensemble members. The vertical lines show the selected fitting time points $t_i$ and the area in between the final $t_s$ and $t_e$ markers is shaded in light blue. For the adaptive time marker scheme we used single hour offsets for the single point results, four hour offsets for dual-point results and eight hours for the triple point result. In all cases the offset was reduced by one hour per iteration of the sequential ABC-SMC algorithm.

The model initialization time $t_0$ is set to a specific date for both events, the initialization distance $r_0$ is fixed to 35 solar radii, and the initial speed $v_0$ and background solar wind speed $v_sw$ are bounded within a smaller range. All other parameters are largely unbounded within a sensible range. For the April event we set $t_0 =$ 2020-04-15T23:00 UTC, $v_0 \in [350, 750]$ and $v_{sw} \in [275, 375]$. For the May event we use $t_0 =$ 2020-05-26T14:00 UTC, $v_0 \in [300, 650]$ and $v_{sw} \in [250, 450]$. The initialization times, distances and CME initial speeds are roughly estimated from the remote observations from STEREO-A. Due to projection effects from the images we can only estimate the lower bound for the speed. The solar wind speed is taken from in situ data prior to the CME arrivals.

For the April event we largely used equidistant fitting points with a one hour separation. While this was also possible for the SolO measurement for the May event this was no longer the case at Bepi or Wind due to the strong perturbations at the center. As can be seen in the Figure \ref{fig:results_insitu} we used an extremely low number of points for these two measurements which are located towards the endpoints of the flux rope signature.

The results for the April event show that we were generally capable of reproducing all three measurements simultaneously with our model. The largest discrepancy can be found within the SolO measurement where we are not able to reproduce the sloped radial component $B_R$ and instead generate an almost constant field as would be expected from a rigid flux rope structure. We also do not find any issues with fitting the varying arrival times which is a strong indication that the general assumptions on the propagation are more or less valid within the vicinity of the three spacecraft. The small jumps that can be seen in the representative ensemble members are artifacts that occur due to our magnetic field model when combined with high aspect-ratios.

\textbf{The results for the May event, despite only being a dual-point fit, are significantly less accurate as can be seen alone from the spread. Since we only took the SolO and Wind measurements into account we can see that the arrival time at Bepi is incorrect with most of the ensemble runs arriving later than expected. For the magnetic field we are more or less able to reproduce the positive constant $B_T$ component although its strength is underestimated in both cases, the bipolar $B_N$ component is overestimated and the radial $B_R$ is far off from the actual measurement at SolO and more or less accurate for Wind.}

\begin{figure*}[h]
\includegraphics[trim=200px 75px 150px 175px, clip, width=.48\linewidth]{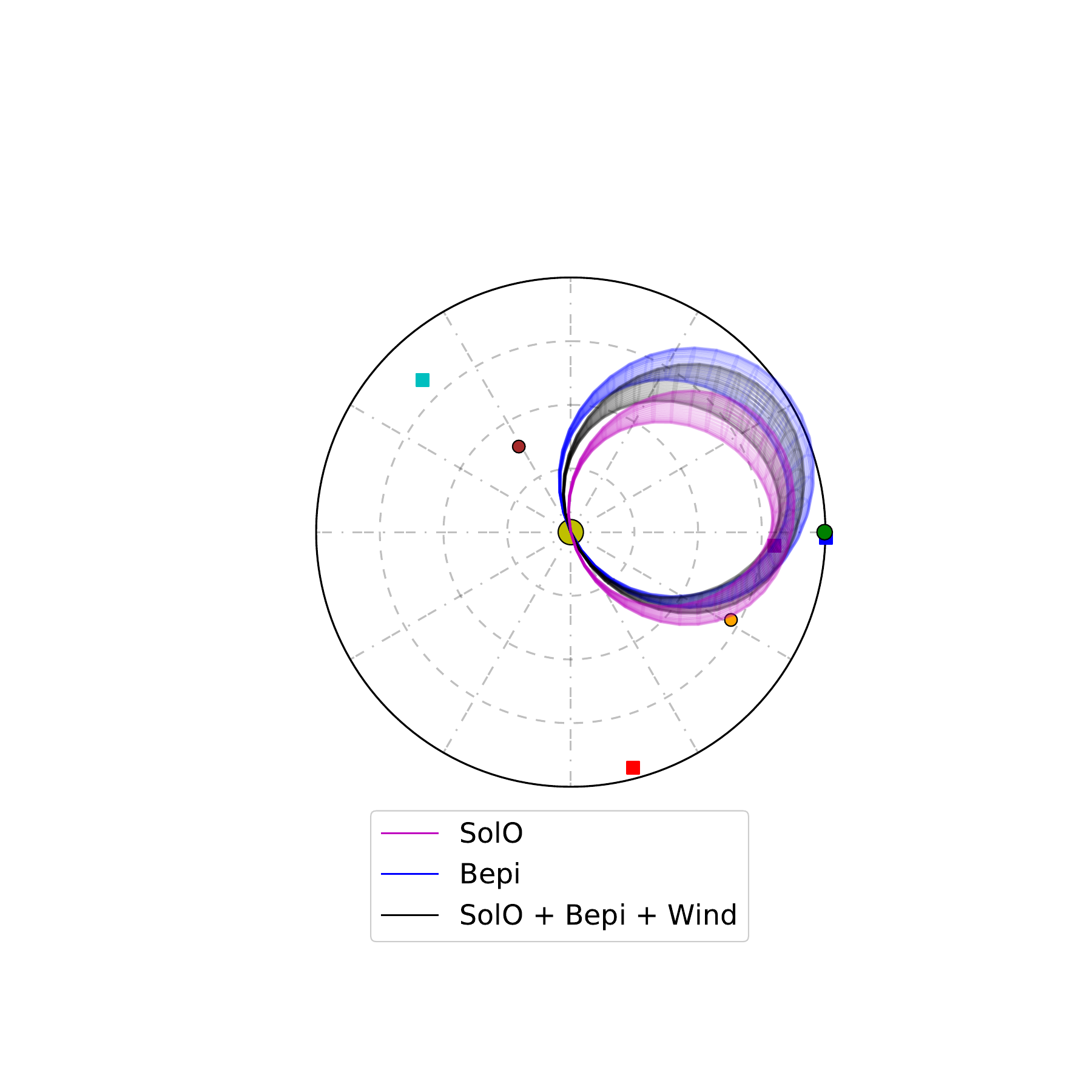}
\includegraphics[trim=200px 75px 150px 175px, clip, width=.48\linewidth]{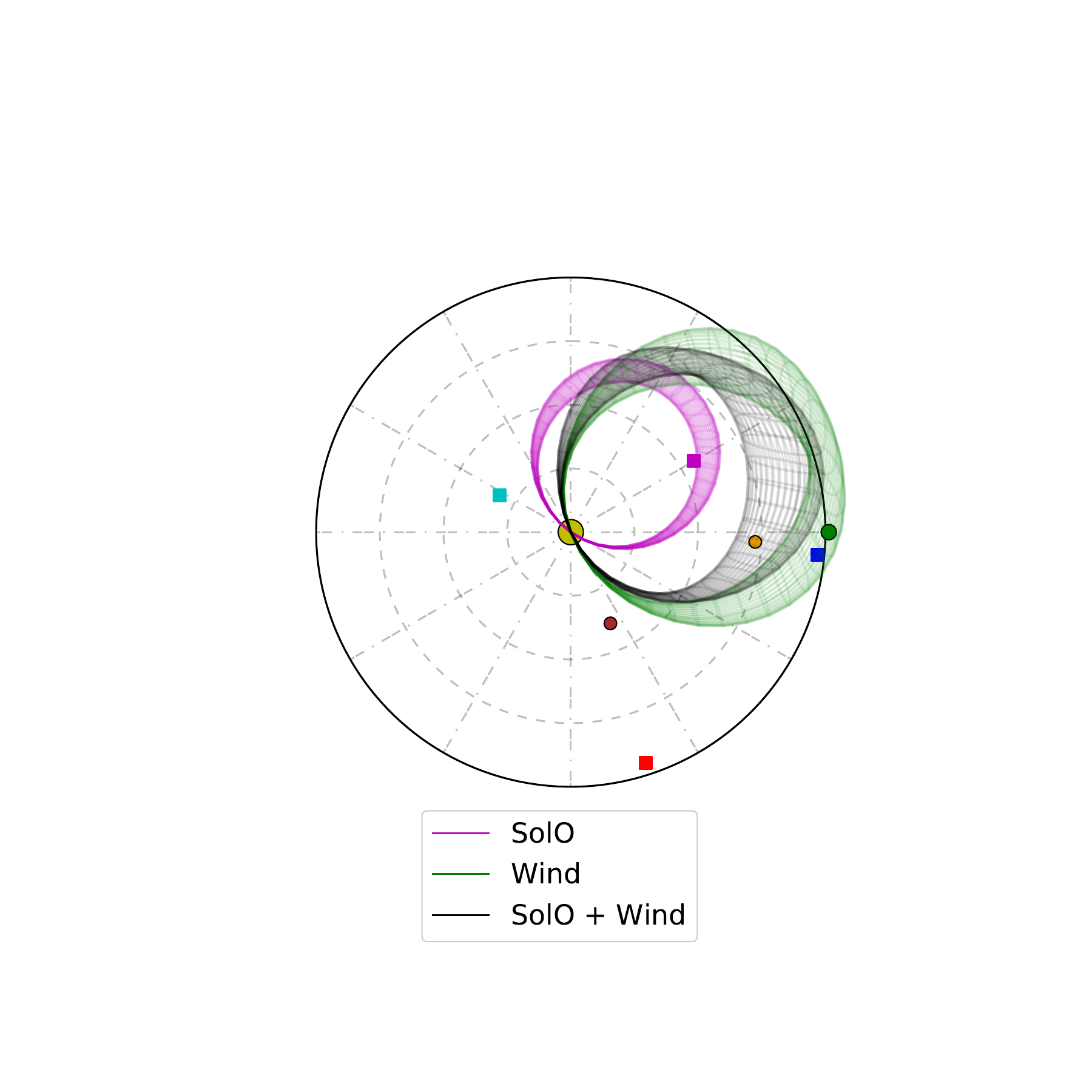}
\includegraphics[trim=200px 325px 150px 275px, clip, width=.48\linewidth]{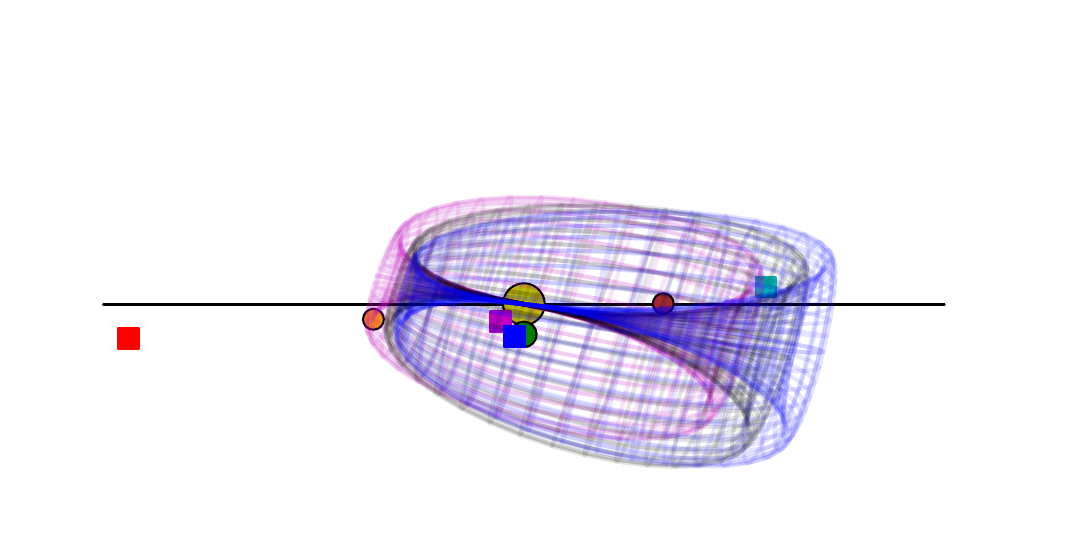}
\includegraphics[trim=200px 325px 150px 275px, clip, width=.48\linewidth]{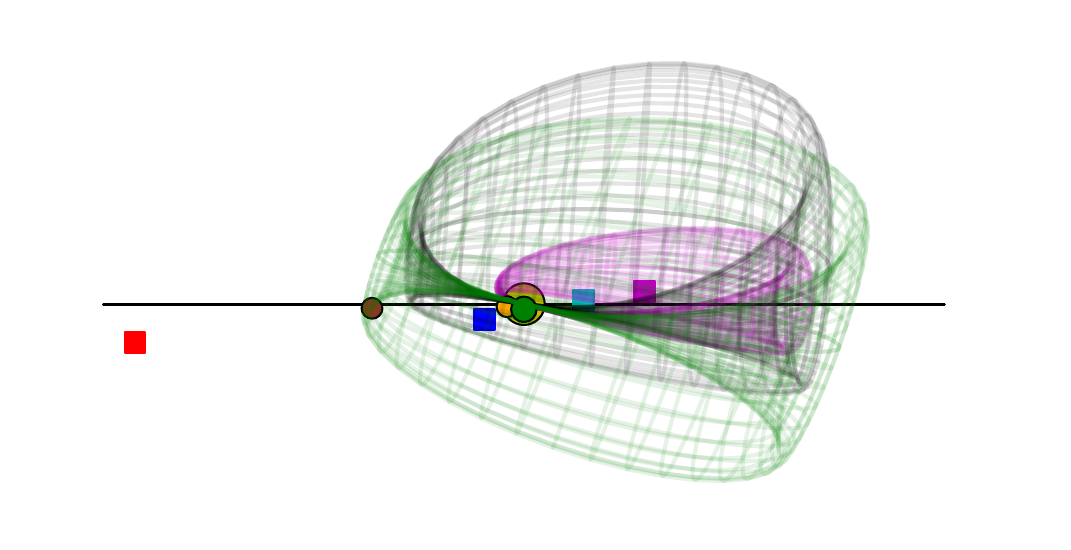}
\caption{\label{fig:reconstructions} Representative 3DCORE geometries for multiple different in situ combinations including the inner planets and satellites using the same color codes as in Figure \ref{fig:events_overview}. The initial parameters for the representative ensemble members are the same as for the representative in situ signatures in Figure \ref{fig:results_insitu}.The two top panels show a view from the top while the bottom panels show a view from the front.}
\end{figure*}

Figure \ref{fig:reconstructions} shows the 3DCORE shapes for representative ensemble members, using the same model parameters as the representative fits in Figure \ref{fig:results_insitu}, from a side/top and a frontal viewpoint. The color codes for the plotted inner planets and spacecraft are the same as in Figure \ref{fig:events_overview}. In both cases we show three different fits, for the April event these are the individual SolO and Wind fits and the combined triple-point fit. In the case for the May event we show the individual SolO and Wind fit including the combined SolO + Wind fit.

\textbf{Table \ref{tab:results_parameters} shows the inferred model parameters for the April and May event using all seven different spacecraft combinations. The final row in each case shows the final achieved error threshold $e^{k}$ for each observation. From these tabular results we can see that the general solutions for the geometry of the April event do not change much when using different observations. A larger difference can be found in the parameters that describe the CME kinematics with the single point observations showing a large range of possible values for $v_0,˛ v_{sw}$ and $\Gamma$. While the $\Gamma$ parameter seems to be largely unconstrained in the multi point results the variations on the initial CME velocity and solar wind background speed are reduced significantly showing that these multi point observations, combined with our rigid CME geometry set stronger constraints for the kinematics.}

\textbf{The ensemble solutions for the May event show larger variations, specifically concerning the orientation and propagation direction of the flux rope . Specifically there is a larger difference in the longitudinal propagation parameter that spans a range of over 35\,deg. Due to these larger variations it is harder to find general solutions that can reproduce the observed measurements for two or three spacecraft, as can be seen in the large error values $\epsilon_\textrm{final}$ for the dual-point fits. These issues clearly show that our assumed geometry and model has significant issues attempting to describe the May event.}

\begin{table*}
\caption{3DCORE parameter fitting results for the April (top) and May (bottom) event in HEEQ coordinates from Solar Orbiter (S), Wind (W) and BepiColombo (B). The $\epsilon_\textrm{final}$ values describes the last $\epsilon^k$ values that were achieved during the iterative fitting algorithm. The shown $\tau$ values were directly computed from the fitted $T_\tau$ (not shown) and $\delta$.}
\label{tab:results_parameters}
\centering   
{\renewcommand{\arraystretch}{1.5}
\begin{tabular}{c r r | c c c | c c c || c}    
\hline\hline                
&& Units  &  SolO &  Bepi &  Wind &  S + B &  S + W &  B + W &  S + B + W\\
\hline
\multirow{11}{*}{\rotatebox[origin=c]{90}{19 April 2020}}&lon&[$deg$]&$15^{+3}_{-3}$&$28^{+3}_{-3}$&$24^{+3}_{-3}$&$25^{+2}_{-2}$&$22^{+2}_{-2}$&$25^{+2}_{-3}$&$24^{+2}_{-2}$\\
&lat&[$deg$]&$-3^{+6}_{-5}$&$-7^{+7}_{-7}$&$-9^{+4}_{-4}$&$-5^{+6}_{-5}$&$-10^{+6}_{-5}$&$-7^{+5}_{-5}$&$-7^{+5}_{-4}$\\
&inc&[$deg$]&$-14^{+3}_{-3}$&$-8^{+3}_{-3}$&$-15^{+3}_{-3}$&$-11^{+2}_{-2}$&$-13^{+3}_{-3}$&$-13^{+3}_{-3}$&$-12^{+2}_{-2}$\\
&$d_{1au}$&[$au$]&$0.10^{+0.010}_{-0.009}$&$0.09^{+0.010}_{-0.009}$&$0.08^{+0.008}_{-0.007}$&$0.10^{+0.006}_{-0.006}$&$0.09^{+0.006}_{-0.005}$&$0.09^{+0.007}_{-0.005}$&$0.09^{+0.005}_{-0.004}$\\
&$\delta$&&$8.27^{+2.55}_{-3.25}$&$8.43^{+2.43}_{-3.16}$&$8.37^{+2.52}_{-3.27}$&$8.73^{+2.23}_{-3.55}$&$7.86^{+2.79}_{-3.60}$&$8.53^{+2.38}_{-3.28}$&$8.84^{+2.19}_{-3.26}$\\
&$v_0$&[$kms^{-1}$]&$320^{+33}_{-13}$&$422^{+104}_{-66}$&$385^{+113}_{-52}$&$323^{+17}_{-15}$&$321^{+16}_{-14}$&$360^{+97}_{-32}$&$319^{+15}_{-13}$\\
&$\tau$&&$-0.55^{+0.20}_{-0.34}$&$-1.02^{+0.34}_{-0.60}$&$-1.40^{+0.37}_{-0.85}$&$-0.75^{+0.23}_{-0.51}$&$-0.81^{+0.28}_{-0.63}$&$-1.15^{+0.34}_{-0.69}$&$-0.87^{+0.22}_{-0.54}$\\
&$b_{1au}$&[$nT$]&$13.44^{+0.60}_{-0.64}$&$14.76^{+0.70}_{-0.66}$&$15.13^{+0.75}_{-0.77}$&$13.99^{+0.58}_{-0.53}$&$13.59^{+0.72}_{-0.70}$&$15.00^{+0.74}_{-0.75}$&$14.16^{+0.56}_{-0.55}$\\
&$\Gamma$&&$1.30^{+0.47}_{-0.61}$&$1.22^{+0.51}_{-0.52}$&$1.29^{+0.46}_{-0.53}$&$1.13^{+0.53}_{-0.49}$&$1.06^{+0.57}_{-0.52}$&$1.29^{+0.45}_{-0.56}$&$1.11^{+0.54}_{-0.49}$\\
&$v_{sw}$&[$kms^{-1}$]&$293^{+27}_{-13}$&$318^{+28}_{-22}$&$307^{+36}_{-18}$&$359^{+10}_{-14}$&$351^{+12}_{-19}$&$317^{+39}_{-22}$&$358^{+9}_{-12}$\\
&$\epsilon_\textrm{final}$&&0.23&0.20&0.20&0.31 + 0.22&0.29 + 0.29&0.23 + 0.22&0.32 + 0.24 + 0.23\\
\hline\hline
\multirow{11}{*}{\rotatebox[origin=c]{90}{29 May 2020}}
&lon&[$deg$]&$20^{+10}_{-15}$&$-15^{+10}_{-9}$&$-13^{+11}_{-10}$&$-16^{+10}_{-3}$&$-8^{+9}_{-6}$&$-23^{+8}_{-5}$\\
&lat&[$deg$]&$4^{+3}_{-4}$&$1^{+12}_{-12}$&$2^{+10}_{-11}$&$6^{+18}_{-13}$&$15^{+11}_{-16}$&$18^{+11}_{-29}$\\
&inc&[$deg$]&$167^{+6}_{-6}$&$164^{+13}_{-14}$&$174^{+13}_{-12}$&$170^{+8}_{-7}$&$169^{+8}_{-8}$&$174^{+10}_{-12}$\\
&$d_{1au}$&[$au$]&$0.13^{+0.010}_{-0.014}$&$0.11^{+0.016}_{-0.016}$&$0.09^{+0.017}_{-0.014}$&$0.13^{+0.012}_{-0.012}$&$0.12^{+0.016}_{-0.016}$&$0.12^{+0.016}_{-0.015}$\\
&$\delta$&&$3.45^{+4.70}_{-1.78}$&$8.51^{+2.38}_{-3.14}$&$8.43^{+2.39}_{-3.17}$&$8.64^{+2.36}_{-2.98}$&$8.75^{+2.22}_{-2.89}$&$10.01^{+1.25}_{-2.16}$\\
&$v_0$&[$kms^{-1}$]&$381^{+104}_{-24}$&$606^{+89}_{-88}$&$540^{+128}_{-104}$&$569^{+116}_{-94}$&$427^{+100}_{-50}$&$651^{+68}_{-83}$\\
&$\tau$&&$2.30^{+3.00}_{-1.18}$&$0.40^{+1.02}_{-0.84}$&$0.66^{+1.21}_{-0.95}$&$2.10^{+1.92}_{-1.43}$&$1.49^{+1.68}_{-1.07}$&$0.31^{+0.72}_{-0.54}$\\
&$b_{1au}$&[$nT$]&$9.58^{+0.59}_{-0.54}$&$9.95^{+2.37}_{-2.27}$&$11.18^{+2.42}_{-2.35}$&$7.05^{+1.45}_{-1.04}$&$7.43^{+1.41}_{-1.08}$&$9.58^{+1.96}_{-2.01}$\\
&$\Gamma$&&$1.71^{+0.21}_{-0.39}$&$0.94^{+0.62}_{-0.50}$&$1.09^{+0.57}_{-0.54}$&$1.05^{+0.58}_{-0.54}$&$1.16^{+0.53}_{-0.57}$&$0.71^{+0.58}_{-0.35}$\\
&$v_{sw}$&[$kms^{-1}$]&$257^{+19}_{-5}$&$410^{+25}_{-37}$&$379^{+43}_{-39}$&$420^{+21}_{-45}$&$396^{+38}_{-37}$&$396^{+38}_{-49}$\\
&$\epsilon_\textrm{final}$&&0.21&0.49&0.38&0.71 + 0.67&0.67 + 0.59&0.68 + 0.57\\
\hline\hline
\end{tabular}}
\end{table*}

\begin{figure}[h]
\includegraphics[trim=0px 0px 0px 0px, clip, width=\linewidth]{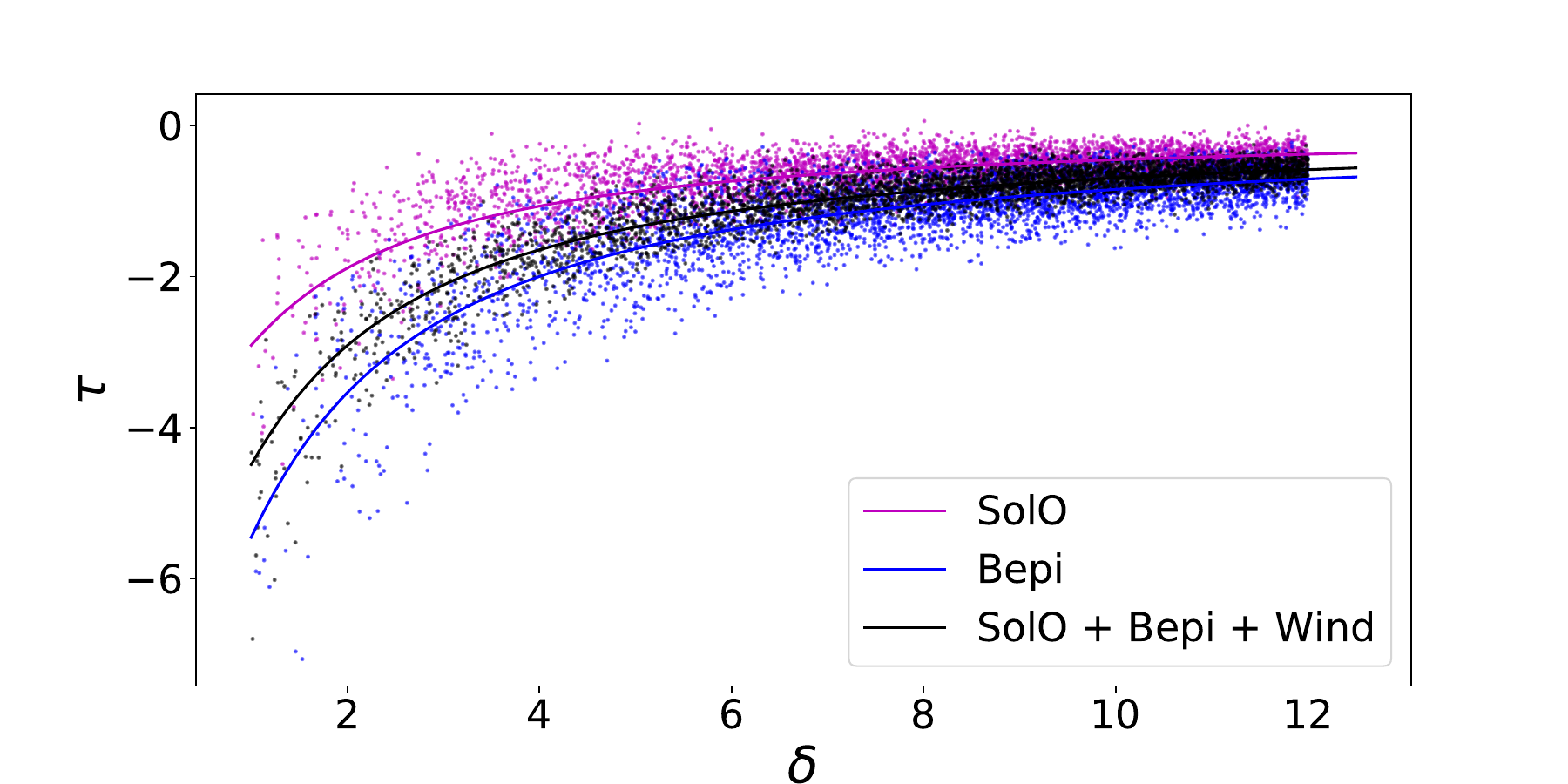}
\caption{\label{fig:degeneracy} Scatter-plot showing the $\tau$-$\delta$ degeneracy for the April event with the SolO, Bepi and Solo+Bepi+Wind ensembles. The solid lines show the $\tau \propto \delta / E(\delta)$ trend.}
\end{figure}

Figure \ref{fig:degeneracy} shows a scatter plot for the twist and aspect-ratio solutions from the overall ensembles of the April event. In this plot we see that the $\delta - \tau$ degeneracy persist despite the high bias towards higher $\delta$ values. As previously explained in Section \ref{sec:bias_corr} this can be explained due to a purely geometrical effect. The solid lines show the $\tau \propto \delta / E(\delta)$ trend for each single case. Unfortunately this means that one of the key issues in W21 remains unresolved. We are not able to solve the $\delta - \tau$ degeneracy using dual or triple point observations. Taking a closer look at the spacecraft positions in the April event in Figure \ref{fig:reconstructions} we are able to see that all three spacecraft are at a similar distance below the inferred flux rope axis. This furthermore reduces the amount of information that can be extracted on the aspect ratio within our magnetic field model.

Lastly we can also compare some of the results to the remote observations shown in Section \ref{sec:event_remote}. We find that in general the inferred directions match the given values from the HELCATS catalog very well. For the April event we find a HEEQ longitude and latitude of $15$ to $28\deg$ and $-3$ to $-10\deg$ where the estimated value is $-2/-6$ from the images. For the May event we infer a HEEQ longitude and latitude of $-23$ to $20\deg$ and $1$ to $18\deg$ where the estimated value is $+36/+11$.

\section{Conclusion} \label{sec:conclusion}

As a summary, we have shown in this paper that it is possible to reconstruct multiple in situ flux rope measurements simultaneously using an empirical CME MFR model. We chose two different scenarios to test our approach, with the first (the April event) consisting of three separate close by measurements and the second (the May event) consisting of three measurements at larger longitudinal separations.

In the first case, with the close by measurements, we do not find any major issues in our attempt to reconstruct the magnetic field measurements using our model. The seven different combinations for reconstruction show more or less the same picture further reinforcing the validity of our description on smaller scales.

The second case illustrates the limits of our modeling approach as we are no longer able to consistently describe all three observations simultaneously. This can be attributed to large longitudinal separations of the spacecraft which allows for a vastly different solar wind environment at the different locations. \textbf{The two measurements for Wind and Bepi are strongly perturbed which may be a hint that they only sample the CME structure at the very edge or have undergone interaction with the solar wind or heliospheric current sheet \citep{Winslow_2016}. Such interactions may impose a limit to which multi-point magnetic field observations can be used for reconstructing the CME structure depending on the specific configuration of the environment.}

The problem of estimating the cross-section aspect ratio and the twist has not been solved by using multi point observations. Due to the constellation of the April event, where the model depicts all the spacecraft to lie below the flux rope axis at similar distances, it is still not entirely clear if this problem will persist for other multi point events or is inherent. The specific magnetic field model that is used in our analysis may prove problematic in this aspect. Unless we are given clear observations of a crossing both below and above the flux rope axis, i.e.\ a positive and negative unipolar $B_r$ signature we believe that we will not be able to resolve the degeneracy of the cross-section aspect ratio and the twist number when using our model. Further studies using different magnetic field models, specifically models that are more physical and include a proper description elliptical cross-sections \citep[e.g.][]{Nieves_Chinchilla_2018}, may resolve this problem or at least reduce the uncertainties.

In general we believe that our approach works well as a first order approximation for multi point observations in close proximity. Due to the different solar wind environments that can be present over longitudinal separations larger than ca. 10 $\deg$ our model will be limited due to not properly being able to describe the arrival times and flux rope orientation at varying positions. The next obvious step would be the construction of more complex and computationally expensive models that include deformations due to the solar wind environment. \textbf{This will invariably require a proper 3D description of the solar wind within at least 1~AU and a CME model with a deformable CME structure}. As our approach is, in principal, valid for any forward based simulation model the entire fitting pipeline can be re-used for this newer class of models. Our current model can also be used as a first step in feeding a more complicated model with initial conditions in order to limit the volume of the parameter space.

The April event represents one of the best multi point observations so far recorded. New missions have been successfully launched within the last years, including BepiColombo, Solar Orbiter and Parker Solar Probe. Expected launches of other missions such as JUICE combined with the rising solar activity of cycle 25 should allow us to observe many additional multi point events within the next decade. These observations will be important for testing our existing approach and future studies in order to develop models that can better describe the global ICME structure.

\begin{acknowledgements}
A.J.W, C.M., R.L.B, M.A.R., T.A., M.B. and J.H. thank the Austrian Science Fund (FWF): P31521-N27, P31659-N27, P31265-N27. We have benefited from the availability of Solar Orbiter, STEREO, Wind, and BepiColombo data, and thus would like to thank the instrument teams, and the SPDF CDAWeb and Solar Orbiter (\url{http://soar.esac.esa.int/soar/}) data archives for their distribution of data. The Solar Orbiter magnetometer was funded by the UK Space Agency (grant ST/T001062/1). This research was supported by funding from the Science and Technology Facilities Council (STFC) studentship ST/N504336/1 (E.D.) and STFC grant ST/S000361/1 (T.H.). D.H., I.R., H.-U. A. were supported by the German Ministerium für Wirtschaft und Energie and the German Zentrum für Luft- und Raumfahrt under contract 50 QW 1501.

The open source code for the 3DCORE model is available at \url{https://github.com/ajefweiss/py3dcore}, and the \textit{heliosat} package to download and process spacecraft data is accessible at \url{https://github.com/ajefweiss/heliosat}.
\end{acknowledgements}


\bibliographystyle{aa}
\bibliography{refs}

\end{document}